\documentclass{amspaper}

\usepackage{natbib}
\usepackage{graphicx}
\usepackage[mathscr]{eucal}
\usepackage{amsmath}

\newcommand{\cA}{\mathscr{A}}

\newcommand\be{\begin{equation}}
\newcommand\ee{\end{equation}}
\newcommand\etal{\textit{et al }}
\newcommand\ve{\mathbf}
\newcommand{\rme}{\mathrm{e}}
\newcommand{\rmd}{\mathrm{d}}

\title{{\bf The capacitance of pristine ice crystals and aggregate snowflakes}}
\author{Christopher David Westbrook\footnote{\textit{Corresponding author address}: Dr. Chris Westbrook, Department of Meteorology, University of Reading, Berkshire, RG6 6BB, UK; email c.d.westbrook@reading.ac.uk}, Robin J. Hogan and Anthony J. Illingworth}
\affiliation{Department of Meteorology, University of Reading, UK.}

\begin {document}
\maketitle[
\begin{abstract}
A new method of accurately calculating the capacitance of realistic ice particles is described: such values are key to accurate estimates of deposition and evaporation (sublimation) rates in numerical weather models. The trajectories of diffusing water molecules are directly sampled, using random `walkers'. By counting how many of these trajectories intersect the surface of the ice particle (which may be any shape) and how many escape outside a spherical boundary far from the particle, the capacitance of a number of model ice particle habits have been estimated, including hexagonal columns and plates, `scalene' columns and plates, bullets, bullet-rosettes, dendrites, and realistic aggregate snowflakes. For ice particles with sharp edges and corners this method is an efficient and straightforward way of solving Laplace's equation for the capacitance. Provided that a large enough number of random walkers are used to sample the particle geometry ($\sim10^4$) the authors expect the calculated capacitances to be accurate to within $\sim1\%$. The capacitance for our modelled aggregate snowflakes ($C/D_{max}=0.25$, normalised by the maximum dimension $D_{max}$) is shown to be in close agreement with recent aircraft measurements of snowflake sublimation rates. This result shows that the capacitance of a sphere ($C/D_{max}=0.5$) which is commonly used in numerical models, overestimates the evaporation rate of snowflakes by a factor of two. 

The effect of vapor `screening' by crystals growing in the vicinity of one another has also been investigated. The results clearly show that neighbouring crystals growing on a filament in cloud chamber experiments can strongly constrict the vapor supply to each other, and the resulting growth rate measurements may  severely underestimate the rate for a single crystal in isolation (by a factor of 3 in our model setup).

\end{abstract}]

\section{Introduction}
The growth and evaporation of ice particles by diffusion of water vapor on to and away from the surface of ice particles are fundamental processes for the development of ice clouds and precipitation.
The density of water vapor $\rho$ around a (stationary) ice particle is governed by Laplace's equation:
\begin{equation}
\nabla^2\rho=0,
\label{laplace}
\end{equation}
under steady-state conditions (Pruppacher and Klett 1997). Typically the separation between ice particles in a cloud is such that each ice particle may be considered in isolation against a background vapor density $\rho_{\infty}$. If the vapor density at the surface $\rho_s$ is assumed to be constant, this leads to the growth rate:
\begin{equation}
\frac{\rmd m}{\rmd t}=D\int_s\nabla\rho\cdot\rmd\ve{s}=4\pi DC(\rho_{\infty}-\rho_s),
\label{gauss}
\end{equation}
by application of Gauss's law over the particle surface $s$, where $D$ is the diffusion coefficient for water vapor in air, and the capacitance $C$ characterizes the shape and size of the ice particle.

The assumption of constant $\rho_s$ implies a constant surface temperature. The temperature $T_s$ at any given part of the surface is determined by both the flux of vapor on to that part of the crystal, resulting in the release of latent heat; and by the rate at which that heat is conducted away from the crystal. Since the equations for diffusion of vapor and transport of heat take the same form, even if a portion of the surface has a larger flux of vapor on to it and becomes hotter through the release of latent heat, it should lose that extra heat at a correspondingly higher rate, resulting in a constant temperature across the whole particle surface. 
In practice the thermal conductivity of ice is approximately two orders of magnitude larger than the thermal conductivity of moist air, so that even if there is some surface migration of the water molecules, the surface temperature should remain almost uniform.

The assumption of constant $\rho_s$,$T_s$ motivates an analogy with electrostatics, where the results for the capacitance $C$ are well known for simple geometries (see table \ref{known}). However, exact solutions for the non-smooth shapes of natural ice particles are not available, and in numerical models (eg. Wilson and Ballard 1999) a capacitance based on one of the shapes in table \ref{known} is usually applied in its place. It is far from clear whether this approximation is a reasonable one, or what the capacitance of realistic ice particles actually is, particularly for large particles which may be complex aggregates. Chiruta and Wang (2003) suggested a refinement to this situation by approximating bullet rosette crystals by a set of smooth lobes and solving equation \ref{laplace} numerically using a finite-element method to obtain the vapor density $\rho$ around the particle, and capacitance $C$. Recently, the same authors have also applied this procedure to solid and hollow columns, estimating the capacitance for five model aspect ratios (Chiruta and Wang 2005). McDonald (1963) and Podzimek (1966) used metal models to simulate realistic ice particles and measure their capacitance; however experimental uncertainties limited the accuracy of such estimates.

\begin{table}
\caption{\label{known}Theoretical capacitances for simple shapes (McDonald 1963) . Note that electrical capacitances (in Farads) are usually normalised as $4\pi\epsilon C$, where $\epsilon$ is the permittivity of the surrounding medium in $\mathrm{Fm}^{-1}$.}
\center{
\begin{tabular}{ll}
\hline\hline
Shape&Capacitance\\
\hline
Sphere, radius $r$&$C=r$\\
Thin disc, radius $r$&$C=2r/\pi$\\
Prolate spheroid: major,&$C=A/\ln[(a+A)/b],$\\
~minor semiaxes $a,b$&~where $A=\sqrt{a^2-b^2}$\\
Oblate spheroid: major,&$C=ae/\sin^{-1}e,$\\
~minor semiaxes $a,c$&~where $e=\sqrt{1-c^2/a^2}$\\
\hline
\end{tabular}}
\end{table}

In this paper a Monte Carlo method for calculating the capacitance of realistic ice particles is described. The trajectories of diffusing water molecules onto the surface of the ice particle are directly sampled using random walks. The fraction of walks which intersect the modelled ice particle provide an estimate for the flux of water molecules onto the particle, and therefore its capacitance. For the non-smooth shapes of realistic ice particles, this sampling approach turns out to be an accurate and efficient method of solving Laplace's equation for $C$. Also, since we are sampling \textit{steady-state} diffusion onto a stationary crystal, statistics for $C$ can be built up sampling one random walk trajectory at a time, and this means that very little computer memory is required. It is interesting to note that random walker sampling has recently been applied in the electrostatics community to calculate electrical capacitances for conductors with sharp edges (eg. Hwang and Mascagni 2004) since it bypasses many of the artefacts introduced by boundary-element, finite-difference and finite-element methods which can cause systematic errors (Wintle 2004).

The capacitance of a number of model pristine ice crystal types is sampled using this method, along with the capacitance of some aggregate snowflakes from the simulations of Westbrook \etal (2004). These shapes are much more realistic than those that are used in numerical models at present, and therefore we should have much more confidence that the results given here accurately represent the capacitance of natural ice particles. This is very important if quantitative comparisons are to be made between observations and model predictions of deposition and evaporation (eg. Forbes and Hogan 2006).

\section{Method}
The simplest Monte Carlo model for solving (\ref{laplace}) runs as follows. A random walker is placed on a large sphere, radius $R_{\infty}$. Its position is chosen at random with uniform probability over the sphere's surface. This represents the far-field where the surfaces of equal vapor density are spherical. Once released from this sphere, the walker is allowed to diffuse around, taking steps much smaller than any characteristic length scales in the target ice particle (say 0.1\% of the smallest length scale for accurate sampling). Each step is taken in a random direction. The walker's motion is tracked until either i) it hits the surface of the particle, or ii) it escapes outside the large sphere $R_{\infty}$ (where the `background' vapor density $\rho_{\infty}$ is fixed). Note that we have treated the ice particle as a perfect absorber with $\rho_s=0$, but from (\ref{gauss}) we see that our calculated capacitances will be valid for any value of $\rho_s$ (including both deposition and evaporation) provided that it is constant over the ice particle surface. The process is repeated for a large number of walkers until an accurate sample of hits and misses has been obtained: then the capacitance may be calculated as $C=f\times R_{\infty}$ where $f$ is the fraction of walkers that hit the ice particle. It is emphasized that the values of the capacitance $C$ estimated using the above method are the same irrespective of the particle surface vapor density/temperature: the capacitance is a function of the shape and size of the particle alone, provided that $\rho_s$ and $T_s$ are constant.

\begin{figure}
\center{\includegraphics[width=3in]{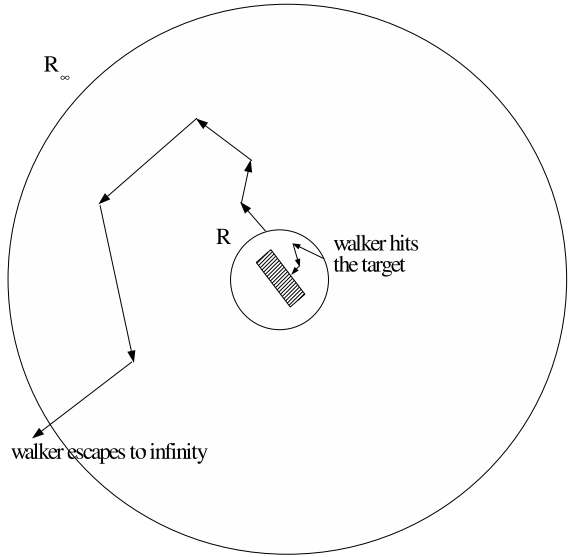}
\caption{\label{walkers}Diagram illustrating the random walker method. Walkers start from the sphere $R$ enclosing the particle, and are tracked along a random walk until they i) hit the particle, or ii) escape beyond a distant spherical outer boundary $R_{\infty}$. Note that outside $R$ the walker is able to take big jumps without any change in its statistics; inside $R$, the walkers can take a jump equal to the distance to the closest point on the surface of the ice particle. In both cases the jumps are in random directions.}}
\end{figure}

In practice, the method outlined above is very inefficient, with a large amount of time being spent tracking the walkers as they take very small steps through large regions of empty space. By making use of the isotropic nature of random walks however, a much more efficient algorithm can be constructed. The first improvement is to start the walkers uniformly on a much closer sphere which just encloses the ice particle (radius $R$ - see figure \ref{walkers}). The distribution of walkers arriving from $R_{\infty}$ onto this closer spherical surface \textit{for the first time} is uniform (because there are no sources or sinks of vapor outside $R$ to distort the spherical symmetry), and therefore starting the walks on this much closer sphere introduces no bias at all on the statistics. The only caveat is that the walkers must still be tracked all the way out to $R_{\infty}$ before they can be assumed to have escaped to infinity. It is important to emphasise that starting the walkers uniformly on $R$ does not force the \textit{total} vapor density to be uniform on $R$ (which would be incorrect). We simply make use of the statistical fact that walkers arriving from $R_{\infty}$ and passing through the surface $R$ for the first time do so randomly and uniformly. The \textit{total} vapor density on $R$ is very much non-uniform, and is distorted by the non-spherical ice particle, which asserts its shape through the statistics of which of the trajectories that pass through $R$ actually hit the particle directly, and which trajectories escape to pass through the surface of $R$ again (and perhaps several times) before hitting the particle or escaping to infinity.

As discussed above, although each walker starts on $R$ it must be tracked back to $R_{\infty}$ before it is assumed to have escaped to infinity (otherwise we would be forcing the vapor density to be uniform on $R$ rather than on $R_{\infty}$). Tracking walkers as they move outside of $R$ is not a great burden however: the walkers may take large steps (as large as the shortest distance back to the sphere $R$) in a random direction since there are no sources or sinks of vapor within that distance and therefore the probability distribution for the new position of the walker is isotropic (and would be even if the walker took a series of very small steps in random directions to travel the same linear distance).

While the random walker is inside $R$ we must be more careful. We can still optimize the step length however, by calculating the minimum distance from the walker's current position to the closest point on the ice particle, and setting this distance to be the length of the next step. The walker is assumed to have hit the ice particle if it comes within some small distance $\delta$ of one of the surfaces. The improvements described above result in a much more efficient algorithm, and importantly no extra approximations have been made. As before, the process is repeated for a large number of walkers so that the particle's geometry is accurately sampled, and the capacitance is given by $C=f\times R$.

\begin{figure}
 \center{\includegraphics[width=3in]{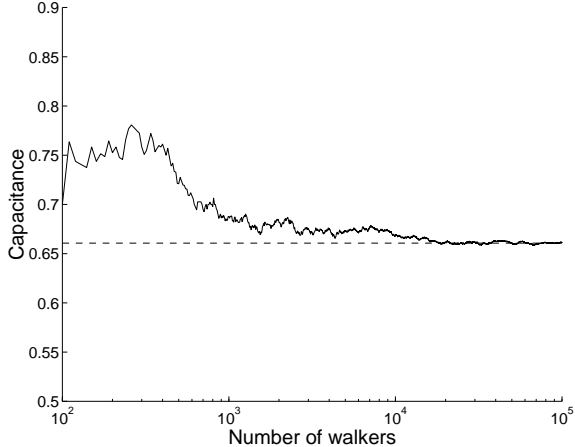}
\caption{\label{cube}Capacitance of unit cube as sampled by an increasing number of random walkers. Dashed line shows the theoretical value of 0.6607 (to 4 d.p.) as calculated by Hwang and Mascagni (2004). The outer boundary was set at $R_{\infty}=500R$, and the thickness of the absorption layer was $\delta=0.001$ (0.1\% of the side length).}}
\end{figure}

The accuracy of the method is determined by three factors: i) the number of walkers used; ii) how far away the `infinite' boundary $R_{\infty}$ is placed; and iii) the thickness of the thin absorbing layer $\delta$. In the calculations presented here the outer boundary is placed at $R_{\infty}=500R$, and the absorption layer thickness $\delta$ is set to be less than 0.1\% of the smallest side length of any of the faces on the particle. There are alternative methods that bypass the construction of such a layer (Mascagni and Hwang 2003); however, for our purposes, simply using a small value of $\delta$ should be sufficient. To test the accuracy of the method, the capacitance of a unit cube has been calculated, which is known to within a tolerance of $10^{-7}$ (Hwang and Mascagni 2004). Provided a large number of walkers are used ($>10^4$), the calculation using the method described above shows excellent agreement to well within 1\% of this value, as shown in figure \ref{cube}. Sensitivity tests indicate that increasing $R_{\infty}$ to a value of $1000R$ has no effect on the estimated capacitance to within $\sim0.1\%$\footnote{We note that it is in fact possible to use the Green's function for a point charge outside a grounded sphere to effectively place the outer boundary at infinity, removing this source of error entirely. This approach also removes the need to track the walkers outside of $R$: see Zhou \etal (1994) for details.}. 

The algorithm has also been tested against particles where the analytic solution is known (see table \ref{known}).
For a thin circular disc with unit radius and thickness of $0.001$, the capacitance is estimated to be $C=0.639$, which matches the exact analytical result for an infinitely thin disc ($C=2/\pi=0.637$) to within 1\%. For a unit sphere we find that our method also agrees with the theoretical value ($C=1$) to well within 1\%. In both cases $10^4$ walkers were used.

\section{Results}
Having established that the method is accurate, we proceed to apply it to a number of model ice particle habits, and the results of this are described below.

\subsection{Hexagonal columns and plates}
\begin{figure}
 \center{\includegraphics[width=2.81in]{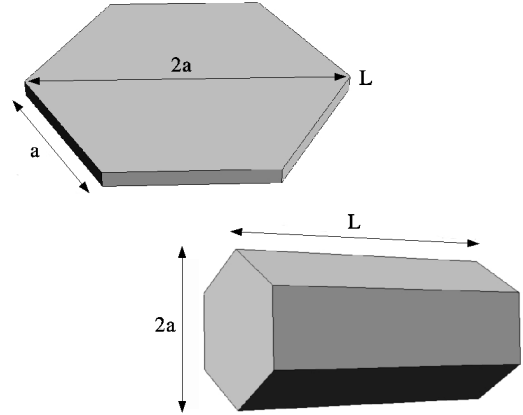}
\caption{\label{habits}Geometry of pristine hexagonal column and plate ice particles. Width $2a$ defined as maximum span across basal (hexagonal) face; length $L$ is the thickness of the plate, or length of the column (ie. the maximum span of the particle in the direction perpendicular to the basal face).}}
\end{figure}

Close to cloud top, one of the most common pristine ice crystal habits is the hexagonal prism (Pruppacher and Klett 1997). Here the random walker method has been used to calculate the capacitance of these shapes, and two examples of this kind of crystal (a column and a plate) are shown in figure \ref{habits}. We define the width $2a$ to be the maximum span across the basal (hexagonal) crystal face, and the length $L$ as the span of the crystal perpendicular to the basal face. The aspect ratio of the crystal is defined as $\cA=L/2a$, ie. columns correspond to $\cA>1$, plates to $\cA<1$.

\begin{figure}[t]
 \center{\includegraphics[width=3in]{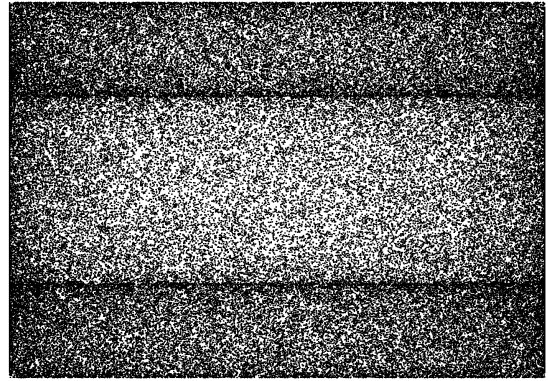}
\caption{\label{hits}Impact positions of random walkers from our simulations onto the surface of a hexagonal column, as viewed in projection from the side. The flux of walkers is highest on the sharp edges and corners of the particle, where the vapor density gradient is largest.}}
\end{figure}

The capacitance of each crystal was sampled by 250,000 walkers, each simulation taking approximately three minutes on a typical desktop PC. The convergence of the sampled capacitances with the number of walkers used was consistent with that for a cube, and as a result we estimate that the results given here for $C$ are accurate to within $\sim1\%$. The distribution of random walkers incident on the faces of a column are shown in side projection in figure \ref{hits}: as one would expect from equation \ref{laplace}, walkers impact on the particle all across the surface, but are strongly concentrated around the particle edges which show up as strong dark lines on the figure, reflecting the high flux of vapor $|\nabla\rho|$ on to these regions. 

\begin{figure}[t]
 \center{\includegraphics[width=3in]{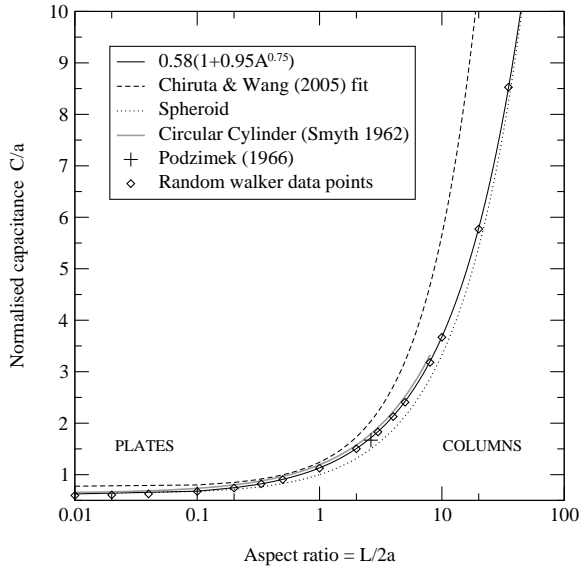}
\caption{\label{col}Capacitance of hexagonal columns and plates with aspect ratios between $\cA=0.01$ (thin plates) and $\cA=50$ (thin columns). Diamonds are data points from our calculations in units of $a$. Solid line is fitted curve $C/a=0.58(1+0.95\cA^{0.75})$. Dashed line shows the fit suggested by Chiruta and Wang (2005). Dotted curve is the capacitance of a spheroid with major and minor axes chosen to match the length $L$ and basal span $2a$ of the crystal. Grey line is the capacitance of a circular cylinder of radius $a$ and length $L$ as calculated by Smyth (1962). Cross indicates the metal-model result from the experiment of Podzimek (1966).}}
\end{figure}

Figure \ref{col} shows the capacitance of hexagonal prisms calculated for a constant width ($a=1$), where the aspect ratio is varied between $0.01$ and $50$.
The equation:
\begin{equation}
C=0.58\left(1+0.95\cA^{0.75}\right)a
\label{ccol}
\end{equation}
closely approximates the data points (to within 1\%) and this curve is overlaid in figure \ref{col} for comparison.

It is common in modelling studies to approximate hexagonal prism type particles as spheroids or circular cylinders. The capacitance for a spheroid with major and minor axes matched to the length and maximal basal span of the hexagonal crystal is shown in figure \ref{col}, and there is reasonable agreement over the range of aspect ratios considered here to within 15\%. Similarly, Smythe (1962) calculated the capacitance of a circular cylinder with diameter $2a_{\mathrm{cyl}}$ and length $L$ as:
\be
C=0.637(1+0.868\cA^{0.76})a_{\mathrm{cyl}}.
\label{cyl}
\ee
It is interesting to note the similar form of equations \ref{ccol} and \ref{cyl}. According to Smythe, (\ref{cyl}) is accurate to within 0.2\% for aspect ratios between $\cA=0$ and $\cA=8$. Comparison shows that the hexagonal prism data lies between the capacitance of a circumscribed cylinder (diameter $2a$, length $L$) and an inscribed cylinder (diameter $\sqrt{3}a$, length $L$), as one would expect since these cases consitute rigid upper and lower bounds for $C$ (see appendix A). The curve for the circumscribed cylinder is plotted in figure \ref{col}.

Also plotted on figure \ref{col} is the fitted curve derived for hexagonal columns and plates by Chiruta and Wang (2005) using a finite-element method. Their linear fit $C=(0.751+0.491\cA)a$ is 10--25\% higher than our data points over the range of aspect ratios which they considered ($\cA=0.2$--$3.33$), indicating that their results are overestimates. Their fit is also higher than that for an enclosing circular cylinder, confirming that their data points are erroneously high (see appendix A). We note that although their capacitances are overestimates, Chiruta and Wang's main conclusion that solid and hollow columns have almost identical capacitances to one another is still likely to be correct, and this is in keeping with metal model measurements (Podzimek 1966). 

The metal model experiments of McDonald (1963) and Podzimek (1966) allowed estimates of the capacitance to be made for hexagonal columns and plates. However there were significant sources of error in these experiments, and the results should be treated with care. Podzimek used an electrolytic tank to estimate the capacitance of a metal hexagonal column with $L=50\mathrm{mm}$ and $2a=19\mathrm{mm}$ (corresponding to $\cA=2.63$), and measured that $C=19.5\mathrm{mm}$, ie. a normalised capacitance of $C/a=2.05$ in apparent agreement with Chiruta and Wang's results. Unfortunately, Podzimek reported a systematic bias of approximately +20\% in his experiments (this was estimated by comparing the measured value for a thin circular disc with the theoretical value from table 1). To resolve this problem, he measured the capacitance for a metal model of a spheroid with the same width and length, and calculated the ratio of the measured column and spheroid capacitances. This ratio ($1.116$) was then assumed to be the same as the ratio of the true capacitances, and using the formula in table 1 he deduced that $C/a=1.67$. This point is plotted on figure \ref{col}, and is in good agreement with our fitted curve ($C/a=1.72$).

Similar experimental difficulties were encountered by McDonald (1963) who placed his metal models inside a walk-in Faraday cage and estimated the capacitance of the arrangement. The connecting lead from the model to the capacitance meter shorted out many of the field lines which should have led from the Faraday cage to the model. This led to $C$ being underestimated by as much as 45\% for a thin circular disc. His resolution of this problem was the same as that of Podzimek, and the ratio of the measured capacitance of the non-smooth models was estimated relative to idealised shapes. From these measured ratios, McDonald estimated that a hexagonal plate has a similar capacitance to that of an equivalent-area circular disc. This is consistent with the data presented here: our fit (\ref{ccol}) predicts the capacitance of a hexagonal plate of zero thickness to be $C=0.58a$, whilst an equal-area circular disc has the almost identical capacitance of $C=0.579a$.

\subsection{`Scalene' columns and plates}
\begin{figure}
 \center{\includegraphics[width=3in]{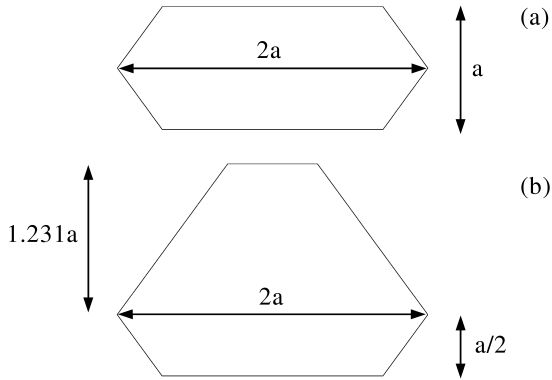}
\caption{\label{distort}Geometry of basal faces for the distorted or `scalene' hexagonal columns and plates. For both types the maximum span across the basal face is defined as $2a$ and all of the internal angles are $120^{\circ}$. The model shapes are: (a) a `flattened' hexagon with 4 short sides and 2 longer sides; (b) a second scalene type with 3 short sides and 3 longer sides.}}
\end{figure}

It has been observed (Bailey and Hallett 2004) that the basal faces of columns and plates are not always perfect regular hexagons, but are often somewhat distorted. Here the capacitance of two so-called `scalene' forms similar to those reported in that paper is calculated, and the impact of breaking the perfect hexagonal symmetry is assessed. The shapes of the model scalene basal faces considered here are shown in figure \ref{distort}. In both cases the maximum span across the basal face is defined as $2a$, and the aspect ratio $\cA$ is defined as before. The capacitances of these scalene crystals have been calculated for three aspect ratios: $\cA=0.1$, $\cA=1$ and $\cA=10$, and the results compared with a perfect hexagonal crystal of the same maximum basal span and length.

The basal face of the first scalene crystal is a `flattened' hexagon, with 4 short sides and 2 longer sides, and a width perpendicular to the maximum basal span of $a$ (compared to $\sqrt{3}a$ for a regular hexagon). The internal angles are all $120^{\circ}$. For $\cA=0.1$, $C$ is reduced by approximately 15\% relative to a regular hexagonal plate; for $\cA=1$ the reduction is $\sim10\%$, and for a column with $\cA=10$ it is only 5\%.

The second model scalene crystal type has 3 short sides and 3 long sides, with a span between opposite sides of $\sqrt{3}a$ (the same as for a regular hexagon). For this type it was found that $C$ is essentially identical to a regular column/plate for all three aspect ratios, to within 3\%. We conclude that distortion of the hexagonal geometry of the crystals has a relatively small impact on their capacitance, and the only significant difference in the growth/evaporation rates is likely to be from surface migration and molecular accomodation effects which are not included here.

\subsection{Bullets and bullet-rosettes}
\begin{figure}
 \center{\includegraphics[width=3in]{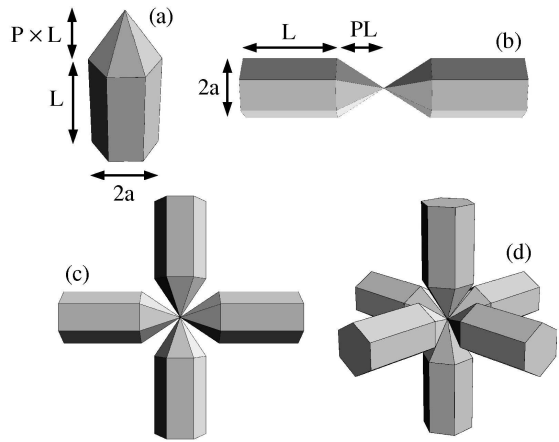}
\caption{\label{bullet}Model geometry for: a) single bullet, b) 2-rosette, c) 4-rosette and d) 6-rosette crystal types.}}
\end{figure}

Bullets and bullet-rosettes are one of the most common crystal habits in cirrus clouds (Heymsfield and Iaquinta 2000). Here we consider some simple model bullets and bullet-rosettes as illustrated in figure \ref{bullet}. The bullets are modelled by a hexagonal column of length $L$, width $2a$, with a hexagon-based pyramid attached to one end. The ratio of the height of the pyramid cap relative to the column length is denoted by $P$, and in what follows we will assume a value of $P=\frac{1}{2}$. The aspect ratio of the columnar section of the bullet $\cA=L/2a$, is defined as before.

\begin{figure}
 \center{\includegraphics[width=3in]{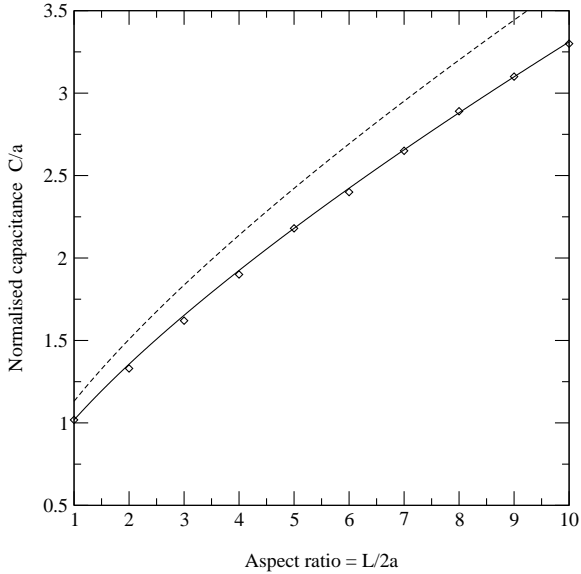}
\caption{\label{cbullet}Capacitance of single bullet crystals as a function of aspect ratio. Diamonds are data points from our calculations in units of $a$. Dashed line shows equation \ref{ccol} for hexagonal columns of the same width and total length as the bullet; solid line is that curve reduced by 10\%.}}
\end{figure}

The capacitance of single bullets with different thicknesses corresponding to aspect ratios $\cA$ between 1 and 10 was calculated. The measured capacitances are plotted in figure \ref{cbullet}. Equation \ref{ccol} for hexagonal columns of the same width ($=2a$) and total length ($=L+PL$) is also plotted on the figure. Comparison between the two shows that our model bullets have a capacitance which is reduced by a uniform 10\% compared to that of a column. Reducing the size of the pyramid cap to $P=\frac{1}{4}$ (following Macke \etal 1996) it was found the capacitance is only 5\% lower than that the complete column, and further reductions in $P$ result in an asymptotic approach to the capacitance of a simple column.

Three model rosettes were constructed, and these are also illustrated in figure \ref{bullet}. The `2-rosette' is simply a linear combination of the two bullets, joined at the tip. Calculations for a range of aspect ratios in the range $\cA=1$ to $10$ indicate that the capacitance of the 2-rosette is reduced by 10--15\% relative to a solid column of the same overall length and breadth, which seems reasonable given its geometry.

\begin{figure}
 \center{\includegraphics[width=3in]{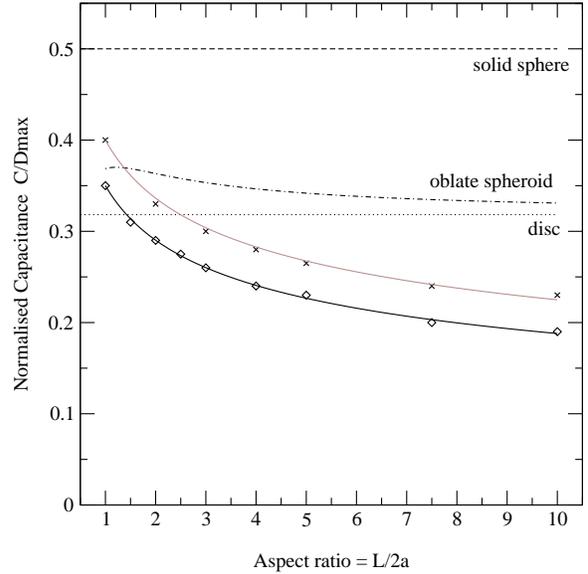}
\caption{\label{crosette}Capacitance of bullet-rosette crystals as a function of the aspect ratio of the arms $\cA=L/2a$. The capacitances are normalised relative to the maximum dimension of the rosette $D_{max}$. Diamonds are the sampled values of $C$ for the 4-arm rosette; crosses are the values for the 6-arm rosette. Solid black and gray lines are the respective fitted curves (see text). The dashed line indicates $C$ for a sphere of the same maximum dimension; the dotted line shows value for a thin disc. The capacitance of an oblate spheroid with a major axis of $2L+2LP$ and minor axis of $2a$ is shown by the dash-dot line.}}
\end{figure}

The `4-rosette' has 4 bullets lying in a plane, each neighbouring bullet separated by $90^{\circ}$. Due to the more complex geometry of this crystal shape, its capacitance has been normalized relative to the maximum dimension $D_{max}$ (rather than the bullet width). Calculations have been performed for a variety of aspect ratios: these values are shown in figure \ref{crosette}. The curve:
\begin{equation}
C=0.35\cA^{-0.27}D_{max},
\end{equation}
was fitted to the data, and appears to provide a close approximation to it. Also shown in figure \ref{crosette} is the capacitance of an oblate spheroid with major and minor axes matched to the overall dimensions of the rosette (major axis $=2L+2LP$, minor axis $=2a$) and this curve overestimates the capacitance quite severely except when the aspect ratio of the bullets is close to unity. The capacitance for a circular disc and a sphere of equal maximum dimension are also shown on the figure for comparison. The value for a disc is a close approximation for $\cA\simeq1.5$, but increasingly overestimates $C$ as $\cA$ increases. The capacitance of the sphere is a strong overestimate, typically a factor of two larger than the rosette.

The final model crystal is a `6-rosette' made of six bullets in a three-dimensional cross shape, each seperated from its neighbours by an angle of $90^{\circ}$. The data for this rosette is also shown in figure \ref{crosette} and the capacitances are approximately 15\% higher than for the 4-bullet rosette. A curve was fitted to this data, and it was found that:
\begin{equation}
C=0.40\cA^{-0.25}D_{max}
\end{equation}
is a good approximation to it. Again, a sphere of the same maximum dimension strongly overestimates the capacitance at all aspect ratios; a disc provides a closer approximation, but still overestimates the capacitance at large aspect ratios (thin arms). 

The results above for four- and six-arm rosettes with $\cA\simeq1.5$ are consistent with the idealised rosettes modelled by Chiruta and Wang (2003). They used a series of smooth lobes to represent each bullet and showed that rosettes with four and six arms had a capacitance broadly similar to that of a circular disc with the same maximum dimension.  An important feature of the new results presented here is that the capacitance of bullet rosettes is sensitive not only to the number of arms, but also to the width of those arms. The aspect ratio of the bullets making up real ice rosettes (as observed by in-situ imaging, eg. Heymsfield \etal 2002) is often rather larger than $\cA=1.5$ and this means that the value of $C$ may in fact be somewhat lower than that of a disc ($\sim20\%$ less for a four-arm rosette with $\cA=3$).

\subsection{Stellar and Dendrite crystals}
\begin{figure}[t]
 \center{\includegraphics[width=3in]{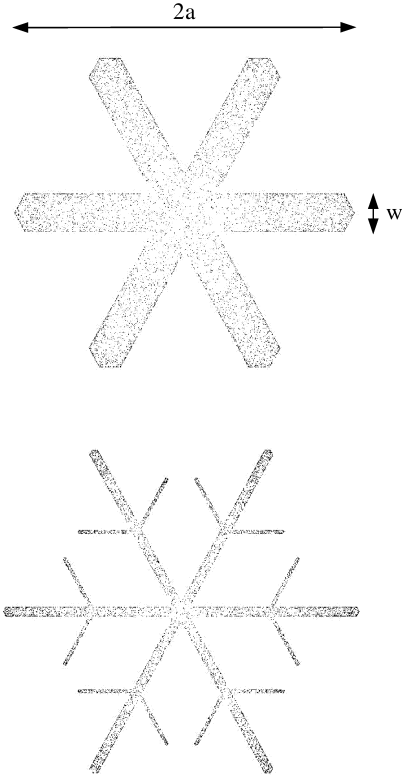}
\caption{\label{dendritehits}Distribution of walkers impacting on the surface of star-type crystal (top) and dendrite with secondary branches (bottom).}}
\end{figure}

Two model stellar/dendrite crystal types were constructed. The first was a simple six-armed star shape, where each arm is a `flattened' hexagonal plate similar to those described in section (3b). Each of these arms is separated from its neighbours by an angle of $60^{\circ}$. The arms overlap at the centre of the crystal, and the internal angles of the arms are all $120^{\circ}$. The span between the tips of two opposite arms is defined as $2a$, and the thickness is fixed as $(2a)/100$. The width of the arms is characterised by the ratio $\cA'=w/a$ which we define as the ratio of the separation between the two longest sides of the hexagon $w$ (ie. the arm width) to the arm length $a$. These dimensions are marked on figure \ref{dendritehits}, which shows the distribution of impacting walkers for a star crystal with $\cA'=0.2$.

\begin{figure}[t]
 \center{\includegraphics[width=3in]{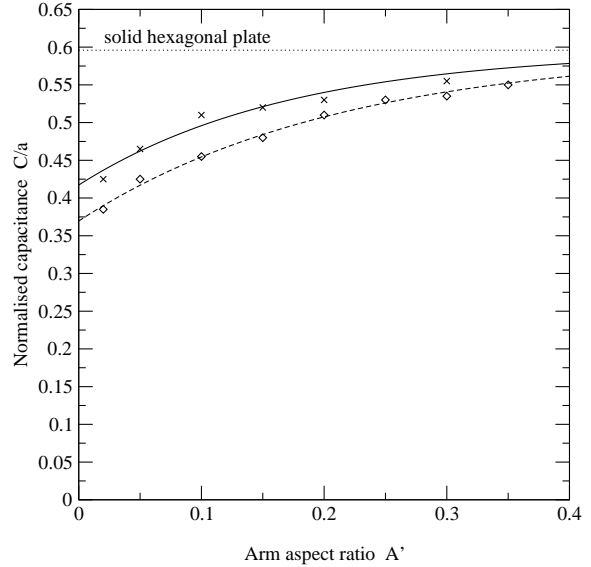}
\caption{\label{dendriteC}Capacitance of stellar/dendrite crystal types. Diamonds are for simple star shapes, crosses are for branched types. The capacitance of a solid hexagonal plate of the same overall dimensions is indicated by the dotted line. Dashed and solid lines are fitted curves for star and branched crystal data respectively (see text).}}
\end{figure}

The capacitance of these star crystals is plotted as a function of the width of the arms in figure \ref{dendriteC}. For a star with thin arms ($\cA'=0.02$) the capacitance is 40\% lower than a hexagonal plate of the same overall dimensions ($=0.596a$ for a plate of thickness $2a/100$). As the arms become thicker, the capacitance rises and asymptotically approaches the value for a solid plate. The curve:
\begin{equation}
C=0.596(1-0.38\rme^{-4.7\cA'})a,
\end{equation}
is also plotted in figure \ref{dendriteC} and approximates the data to within a few percent.

The second model is an adaptation of the star model above, where a pair of secondary branches has been added to each arm, each with one third the dimensions of the main arms. These secondary branches are positioned with one end at the centre of the main arm, and oriented at an angle of $60^{\circ}$ on either side of it. Figure \ref{dendritehits} also shows the distribution of impacting walkers on this crystal type, with $\cA'=0.05$.

The capacitance of this model crystal was calculated for different arm thicknesses, and the data is shown alongside the results for the star shapes in figure \ref{dendriteC}. Adding the branches increases the capacitance somewhat for dendrites with thin arms (for $\cA'=0.05$ the branched dendrite has a capacitance $\sim15\%$ larger than a star). As the branches become thicker they approach the value for a solid plate. The curve:
\begin{equation}
C=0.596(1-0.30\rme^{-5.8\cA'})a,
\end{equation}
is a reasonable fit to the data, and this function approaches the solid plate limit somewhat faster than the simple star crystal.

\subsection{Aggregates}
In this section we consider aggregates of the above ice crystal types. Aggregates are often the dominant particle habit in deep non-precipitating cirrus clouds (Field and Heymsfield 2003, Westbrook \etal 2006), as well as in snowstorms (Jiusto and Weickmann 1973). A recent theoretical model of ice crystal aggregation (Westbrook \etal 2004) has allowed us to produce large samples of realistic `synthetic' ice aggregates. We have calculated the capacitance of these synthetic snowflakes in the expectation that the results ought to be a good approximation to the capacitance of natural ice aggregates.

\begin{figure}[t]
 \center{\includegraphics[width=3in]{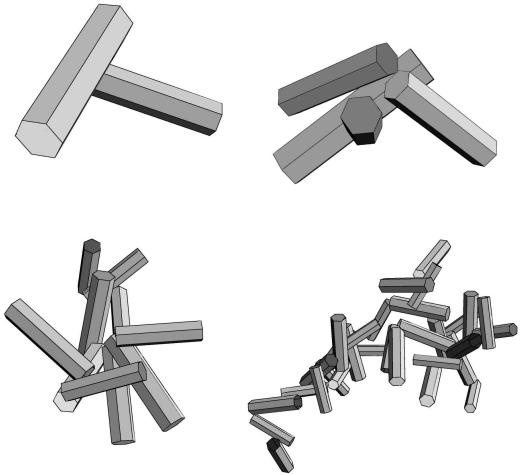}
\caption{\label{aggs}Examples of aggregates made up of 2,4,10 and 32 hexagonal columns, sampled from the simulations of Westbrook \etal (2004). The aspect ratio of the columns is $\cA=4$.}}
\end{figure}

The random walker method was applied to $\sim1000$ synthetic aggregates sampled from the simulations of Westbrook \etal (2004): a few examples are shown in figure \ref{aggs}. We have calculated the average capacitance of the synthetic aggregates, using $10^3$ walkers to sample each individual aggregate. We note that this is a smaller number of walkers than was used for the pristine ice crystal habits; however, the computational cost is increased for complex shapes, and we are averaging the results over many realisations of the aggregate geometry (every aggregate snowflake is different), so we expect that our eventual statistics should be accurate. In any case the results for the unit cube in section 2 indicate that the error in the calculated values for each individual aggregate should be less than 10\%. 

\begin{figure}[t]
 \center{\includegraphics[width=3in]{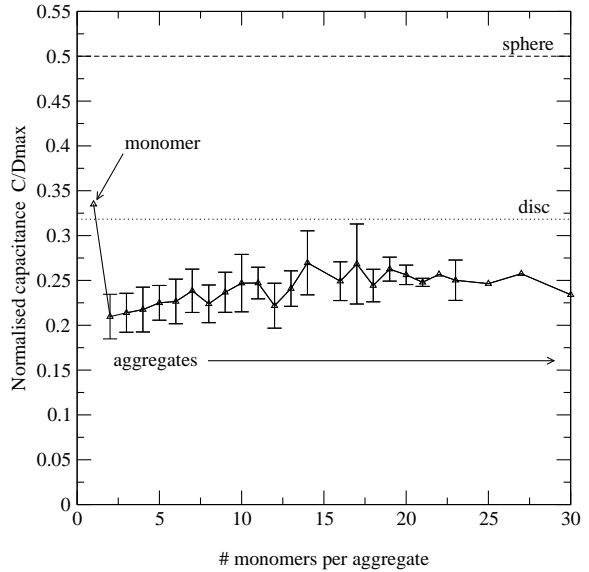}
\caption{\label{aggcol}Capacitance of aggregate snowflakes as sampled from the simulations of Westbrook \etal (2004). The monomer crystals were hexagonal columns with $\cA=2$. The calculated values of $C/D_{max}$ were binned and averaged as a function of the number of crystals per aggregate, as shown on the horizontal axis. Error bars are one standard deviation, points with no error bar indicate the value for a single aggregate. The normalised capacitance asymptotically approaches a value of $C/D_{max}\simeq0.25$. Overlaid are theoretical values for a sphere (dashed line) and a thin disc (dotted line) with the same maximum dimension.}}
\end{figure}

For each aggregate, the ratio of capacitance to maximum dimension $C/D_{max}$ was calculated, and the results binned and averaged as a function of the number of crystals in the aggregate: this is shown in figure \ref{aggcol}. The `monomer' crystals in this case were hexagonal columns with an aspect ratio of $\cA=2$. The capacitance of the monomer columns is $C/D_{max}=0.34$ (see section 3a), but for an aggregate of just two columns this is reduced to $0.21\pm0.02$, since two columns stuck together is a much more open geometry than a single column. As more columns are aggregated the normalised capacitance rises to an asymptotic limit of $0.25\pm0.02$. This value is in strong agreement with recent in-situ aircraft measurements of aggregate snowflake sublimation made by Field \etal (2007) during a Lagrangian descent through a subsaturated portion of an ice cloud, where a value of $C/D_{max}=0.26$ was estimated.

\begin{figure}
\center{\includegraphics[width=3in]{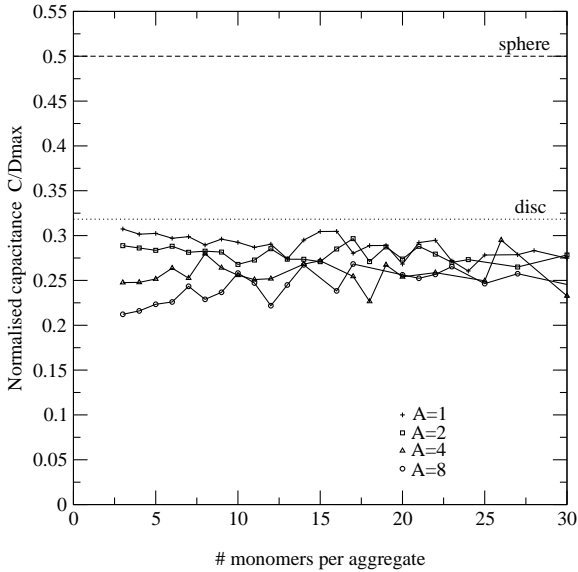}
\caption{\label{aggs_comb}
The capacitance of aggregates with different monomer ratios $\cA$ between 1 and 8. All of the curves approach an asymptotic value in the range 0.25--0.28.}}
\end{figure}

The calculations described above were repeated for different monomer aspect ratios, and this is shown in figure \ref{aggs_comb}. We find that for all the monomer shapes considered the normalised capacitance approaches an asymptotic value of between 0.25 (thin columns $\cA=8$) and 0.28 (squat columns $\cA=1$). The area ratio of these aggregates (the ratio of the particle's projected area to the area of a circle of diameter $D_{max}$) was measured and was also found to approach an asymptotic value of between 0.1 ($\cA=8$) and 0.35 ($\cA=1$). 

Changing the monomer crystal type to bullet-rosettes rather than columns is found to have very little effect on the capacitance, again yielding asymptotic values in the same range. This capacitance is smaller than a disc or sphere of the same maximum dimension, and the implication is that numerical models which assume these simple shapes are overestimating the growth/evaporation rate, by a factor of two in the case of the sphere.

It may perhaps appear counter-intuitive that the capacitance of aggregates should approach a constant value relative to their maximum dimension ($C/D_{max}=\mathrm{constant}$). The structure of the aggregates approaches a fractal geometry (Westbrook \etal 2004) and as a result becomes increasingly open as the aggregates grow to contain more and more ice crystals. Because of this, the effective density decreases with size as $(D_{max})^{-1}$ (in agreement with experimental data: Brown and Francis 1995, Heymsfield \etal 2002). One might therefore anticipate that $C/D_{max}$ would be reduced for large aggregates, since there is more empty space within a radius $D_{max}/2$ of the particle centre, and therefore (one would imagine) more opportunity for a water molecule coming within that radius to escape.
However, the results of Ball and Witten (1984) show that water molecules following a Brownian path are exceptionally efficient at exploring three dimensional space, and because of this the ice particle appears essentially opaque to the incident water molecules (i.e. a fixed fraction of those venturing within $D_{max}/2$ will be absorbed). As a result, $C/D_{max}=\mathrm{constant}$ is in fact the physically sensible result for aggregates (see appendix B for further details).

\section{Screening}
Growth rates derived from laboratory experiments such as those of Bailey and Hallett (2004) are usually measured from crystals which grow not in isolation, but surrounded by other growing crystals. Screening of one ice crystal by another may be an important effect in light of the above discussion on aggregates, since the diffusing water molecules are very efficient at exploring the space around the particle, so neighbouring crystals may constrict the vapor supply to one another. An understanding of such screening effects is important if laboratory data are to be accurately interpreted.

\begin{figure}[t]
 \center{\includegraphics[width=2.5in]{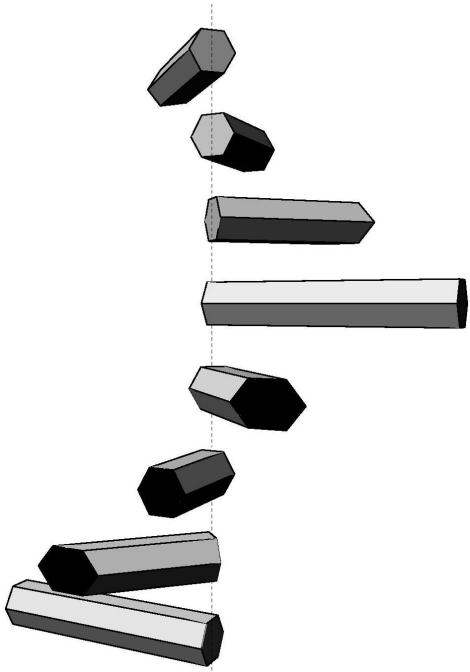}
\caption{\label{screen}Model `laboratory' setup of 8 identical hexagonal column crystals. Each is rotated by an angle of $45^{\circ}$ about the vertical axis relative to its neighbours, and offset in the vertical direction by a crystal's width, in an attempt to model crystals growing on a vertical filament (indicated by the dotted line). The growth rate of the innermost crystals is approximately one third the growth rate of a column growing in isolation.}}
\end{figure}

For our model setup we consider 8 identical hexagonal columns with $\cA=5$ in a `spiral staircase' geometry, growing with their longest axis in the horizontal direction. Neighboring crystals are offset in the vertical direction by a column's width and are rotated around the vertical axis by $45^{\circ}$. This is shown in figure \ref{screen} and is roughly modelled after photographs of the experimental setup in Bailey and Hallett (2004). The surface vapor density and temperature are assumed to be the same on each crystal, and $10^4$ walkers were used to sample the capacitance of the crystals. It was found that the vapor supply to the innermost crystals is inhibited to such an extent that their capacitance is reduced to approximately one third of its value for a single column in isolation ($C/a\simeq0.75$ compared to $2.41$ in isolation). This result demonstrates how sensitive the vapor field around a crystal is to other sources/sinks of vapor in the vicinity. It also shows that if the growth rates from experiments are to be compared to theoretical capacitances, the crysals must be grown in isolation, or separated from one another as much as possible. Electrodynamic trapping techniques (eg. Swanson \etal 1999) where ice particles are levitated in an electric field and allowed to grow in isolation may be useful in this respect, although to the authors' knowledge the results have so far been limited to small particles, less than $100\mu$m in size.

\section{Discussion}
The capacitance for pristine ice crystals and aggregate snowflakes has been calculated using a new Monte Carlo method, and these new values should be an improvement on the traditional approximations of smooth spheroids or discs. Application of these capacitances to estimate the actual growth/evaporation rates requires a prescription for the difference between the surface and the far-field vapor densities $(\rho_s-\rho_{\infty})$. Using the Clausius-Clapeyron equation, an expression for $\rmd m/\rmd t$ in terms of the supersaturation with respect to ice $(S-1)$ is obtained:
\begin{equation}
\frac{\rmd m}{\rmd t}=4\pi C\times\frac{S-1}{A+B}
\label{ab}
\end{equation}
where the parameters $A(T)$ and $B(T,P)$ are given in Pruppacher and Klett (1997) and depend on the ambient temperature $T$ and pressure $P$. We believe that the results described in this paper may help to improve the estimation of growth and evaporation rates calculated in numerical weather prediction and cloud resolving models, especially for spatially extended particles such as bullet rosettes, dendrites and aggregates, where modelling the particle as a simple sphere of the same maximum dimension (eg. Liu \etal 2003, Khain and Sednev 1996) can overestimate $|\rmd m/\rmd t|$ by a factor of 2. This work may be particularly valuable to new precipitation models which predict ice particle habit from the model temperature and humidity (eg. Woods \etal 2006), allowing an appropriate capacitance to be applied for each predicted habit. We note that models which assume the capacitance of an equivalent \textit{volume} sphere can lead to errors in $C$ which are size dependent, depending on the mass-dimension relationship assumed. Using equivalent volume spheres is an unphysical approach to estimating the capacitance, since $C$ is determined by the physical dimensions of the particle and not its volume\footnote{We note that the capacitance of an equal volume sphere is always an underestimate for the capacitance of any non-spherical shape (P\'{o}lya and Szeg\~{o} 1951).}. Parameterising $C$ in terms of $D_{max}$ is a more natural approach since it represents the overall dimension of the particle and is also the parameter usually estimated from aircraft observations. 

The key limitation of the present study is the question of whether the growth and evaporation rates of ice crystals are simply dominated by the rate at which vapor impinges on/diffuses away from the ice crystal surface, or whether surface effects plays a significant role in limiting the growth/evaporation rates. For evaporation of ice particles, Nelson (1998) showed that the mass loss rate is controlled by (\ref{ab}) and that surface migration of water molecules can be neglected, although molecular accomodation may play a role for very small, cold crystals (Magee \etal 2006). This conclusion appears to be supported by in-situ and laboratory imaging of sublimating ice particles (Korolev and Isaac 2004, Swanson 1999) which indicate that the vapor flux away from the particles is concentrated at the corners and edges. This in keeping with the expectation for evaporation through bulk diffusion of vapor as described by equation \ref{laplace} without any surface migration. We therefore expect that equation \ref{ab} should be directly applicable to the problem of calculating the evaporation rates of most natural ice particles using the estimates for $C$ presented here.

For deposition the situation is less clear cut: the appearance of the wide variety of different crystal habits at different temperatures and supersaturation levels indicates the influence of molecular accomodation and surface migration on the growth. Despite this, the evidence from a number of experiments is that equation \ref{ab} is a reasonable approximation for the growth rate in many cases, although there is still much uncertainty as to the influence of molecular accomodation on growth rate at different temperatures and crystal sizes as summarised in Magee \etal (2006), Fukuta and Takahashi (1999) and Pruppacher and Klett (1997); in particular, poor accomodation of impinging vapour may be rather important for small, cold crystals. For frozen drops, laboratory measurements (Korolev \etal 2004) have shown good agreement with (\ref{ab}). The new capacitance results presented in this paper should allow a more accurate comparison to be made between theory and experimental data than has previously been possible, and may help experimentalists to estimate the `accomodation coefficient' of complex ice crystals. 

Bailey and Hallett (2004) compared the growth rates of laboratory-grown plate and column crystals grown over a temperature range of $-20$ to $-70^{\circ}$C to the growth rate calculated using (\ref{ab}), modelling their lab crystals as spheroids (which our results indicate should be a reasonable approximation to within 15\%). Their results indicated that the predicted growth rates were only consistent with the measured values for aspect ratios close to $\cA\simeq1$: for more extreme aspect ratios the theoretical values were as much as a factor of four too large for thin columns, and a factor of eight too large for thin plates. This may imply that the water molecules cannot be easily accommodated at these low temperatures. Bailey and Hallett argue that because of the discrepancy between predicted and measured growth rates, the whole `electrostatic' approach is unsuitable for calculating deposition rates at cold temperatures, and they recommend that laboratory-measured growth rates should be used instead. However, the sensitivity of (\ref{laplace}) to screening as discussed in section 4 show that their measured growth rates may be strongly affected by neighbouring crystals, restricting the vapor supply, and reducing the measured capacitance significantly (a factor of 3 in our model setup). This may explain, at least in part, why their estimated growth rates are so much lower than equation \ref{ab} predicts. The supporting glass filament may also play a screening role, particularly for thin columns where the filament thickness ($50$--$70\mu$m) is the same size or larger than the basal faces of the crystals. On the other hand, the thermal conductivity of this filament is much higher than the surrounding air, enhancing heat conduction and increasing the growth rate, leading to further uncertainties. The interpretation of laboratory measured growth rates, and their use in evaluating the accuracy of theoretical models therefore requires a great deal of care, and this is an issue where more experiments and theoretical work are urgently required to inform such comparisons. 

A key limitation of capacitance theory is that it is unable to make any prediction about particle habit, and a particular shape must be assumed \textit{a priori}. Growth under equation \ref{laplace} with the (moving) boundary condition $\rho=\mathrm{constant}$ at the surface is a much studied problem in theoretical physics: if the water molecules are simply deposited according to the distribution of vapor flux over the particle surface (eg. figure \ref{hits}) the growth is unstable (Langer 1980, Mullins and Sekerka 1963) and fern-like fractal patterns emerge (eg. Witten and Sander 1981, Bowler and Ball 2005; for a general review see Sander 2000). That ice crystals grow in a much more controlled way indicates the influence that surface diffusion and the anisotropic molecular accommodation of the impinging water vapor have on the particle growth. It seems likely that dendrite crystals are a case in point, with their fern-like structure indicative of unstable growth, and the broad six-fold symmetry indicating the influence of the underlying crystalline anisotropy. This has recently been modelled using a crude anisotropy parameter by Goold \etal (2005) to produce fractal dendrites with hexagonal symmetry. Progress in producing theoretical models which predict habit may well follow this approach, using quantitative data for the anisotropic molecular accomodation coefficient and surface migration effects at different temperatures and supersaturations.

The capacitance of realistic ice aggregates has been calculated for the first time (to the authors' knowledge), and the results are in close agreement with those estimated by Field \etal (2007) from in-situ aircraft observations of evaporating ice aggregates, again indicating that (\ref{ab}) is a good approximation for sublimating ice particles. It should be noted that the correction for the enhancement of the evaporation due to ventilation was estimated using formulas for idealised shapes (Hall and Pruppacher 1976) and this introduces some uncertainty into the comparison. However, since a more accurate estimation of the ventilation effect would require a detailed treatment of the flow pattern around the aggregates, this value is at present the most accurate estimate of $C$ that it is currently possible to make from aircraft observations.

The authors believe that the capacitances presented here are, in any case, the best estimates currently available for the growth and evaporation of realistic ice crystals and snowflakes within the framework of the electrostatic analogy.

\section*{Appendix A}
Here we show that the capacitance of a hexagonal column with width $2a$ and length $L$ (see figure \ref{habits}) must be less than that of a circular cylinder of diameter $2a$ and length $L$. 

Let the surface vapor density be $\rho_s=0$ on the surface of the hexagonal column. Consider the flux of vapor through a cylindrical surface $S_{\mathrm{cyl}}$ with the above dimensions, just enclosing the column. The net flux of vapor $\Phi_{\mathrm{net}}$ through this surface must be the same as that onto the surface of the column since there are no other sinks of vapor present within the enclosed volume. 

\begin{figure}
 \center{\includegraphics[width=3in]{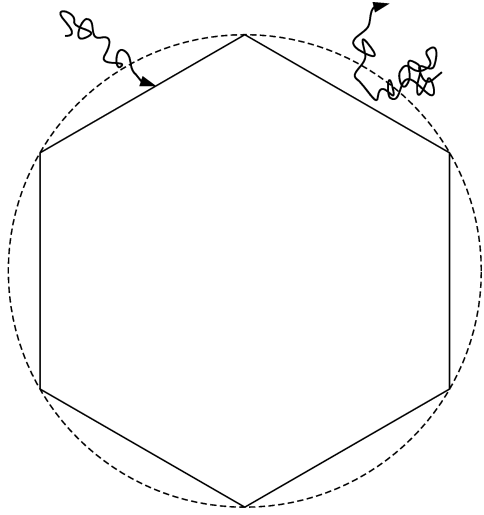}
\caption{\label{blah}Cylindrical surface (dashed line) just enclosing a hexagonal prism (solid line), viewed in projection from one end. A trajectory which passes through the cylindrical surface and is absorbed on the hexagonal prism is illustrated; also shown is a trajectory which passes through the cylindrical surface but escapes again and is not absorbed.}}
\end{figure}

As figure \ref{blah} illustrates, the flux can be split into two components: $\Phi_{\mathrm{net}}=\Phi_{\mathrm{cyl}}-\Phi_{\mathrm{escaped}}$. The flux of water molecules incident from the outer boundary onto $S_{\mathrm{cyl}}$ for the first time is simply given by the flux onto a perfectly absorbing cylinder of the same dimensions $\Phi_{\mathrm{cyl}}=4\pi D\rho_{\infty}C_{\mathrm{cyl}}$ (since this corresponds to every water molecule trajectory being terminated at the point it first intersects $S_{\mathrm{cyl}}$). The value of $C_{\mathrm{cyl}}$ is given by equation \ref{cyl}. Subtracted from this is the flux of water molecules which pass through $S_{\mathrm{cyl}}$ but are not absorbed by the hexagonal prism and escape to infinity $=\Phi_{\mathrm{escaped}}$. Since this latter flux is finite, $\Phi_{\mathrm{net}}<\Phi_{\mathrm{cyl}}$, and by Gauss's law $C<C_{\mathrm{cyl}}$, ie. the capacitance of a hexagonal prism is lower than that of an enclosing circular cylinder.
By the same argument it follows that a circular cylinder of diameter $\sqrt{3}a$ must have a lower capacitance than the hexagonal prism which it inscribes. This allows the construction of upper and lower bounds on the capacitance of a hexagonal prism using the results of Smythe (1962). The violation of the upper bound by Chiruta and Wang's (2005) results shows that their data are overestimates.

\section*{Appendix B}
Here we show that $C/D_{max}$ is a function only of the shape of an ice aggregate, and not its size.
Imagine that the particle is removed and it is replaced with a `ghost' particle which is completely transparent to the diffusing water molecules. The average water molecule follows a Brownian trajectory, tracing a fractal path with dimension $d_w=2$ (Falconer 2003). This means that whilst the water molecule is within a distance $D_{max}/2$ of the ghost particle centre, the resulting path fills a volume $\propto(D_{max})^{d_w}$.
The ghost particle on the other hand, occupies a volume $\propto(D_{max})^{d_i}$, with $d_i\simeq2$ for our aggregates.

Since $d_w+d_i>3$, the points where the particle and the uninterrupted random walk intersect with one another occupy a volume $\propto(D_{max})^3$. Because of this, the aggregates appear essentially opaque to the diffusing water molecules, in the sense that for a given monomer type, a fixed fraction of the molecules which venture within a radius $D_{max}/2$ will come into contact with the ice particle. The ratio $C/D_{max}$ simply represents this fraction, and is therefore independent of size.

\section*{Acknowledgements}
This work was funded by the Natural and Environmental Research Council (grant NER/Z/2003/00643). CDW acknowledges helpful discussions with Mihai Chiruta, Paul Field (NCAR) and Richard Forbes (Met Office).


\begin{thebibliography}{2}
\small

\bibitem[Bailey and Hallett (2004)]{bandh} Bailey, M. and J. Hallett 2004: Growth rates and habits of ice crystals between $-20^{\circ}$C and $-70^{\circ}$C. {\em J. Atmos. Sci.,}
{\bf 61,} 514--544

\bibitem[Ball and Witten (1984)]{ball} Ball, R. C. and T. A. Witten 1984: Causality bound on the density of aggregates. {\em Phys. Rev. A,}
{\bf 29,} 2966--2967

\bibitem[Bowler and Ball (2005)]{bowlerandball} Bowler, N. E. and R. C. Ball 2005: Off-lattice noise reduced diffusion-limited aggregation in three dimensions. {\em Phys. Rev. E,}
{\bf 71,} 011403

\bibitem[Brown and Francis (1995)]{brown} Brown, P. R. A. and P. N. Francis 1995: Improved measurements of the ice water content of cirrus using an evaporative technique. {\em J. Atmos. \& Ocean. Tech.,}
{\bf 10,} 579--590

\bibitem[Chiruta and Wang (2003)]{chiruta} Chiruta, M. and P. K. Wang 2003: The capacitance of rosette ice crystals. {\em J. Atmos. Sci.,}
{\bf 60,} 836--846

\bibitem[Chiruta and Wang (2005)]{chiruta2} Chiruta, M. and P. K. Wang 2005: The capacitance of solid and hollow hexagonal ice columns. {\em Geophys. Res. Lett.,}
{\bf 32,} L05803

\bibitem[Falconer (2003)]{falconer}
Falconer, K. 2003,
\textit{Fractal Geometry: Mathematical Foundations and Applications, 2nd Edition}, John Wiley \& sons, London.

\bibitem[Field and Heymsfield (2003)]{fandh} Field, P. R. and A. J. Heymsfield 2003: Aggregation and scaling of ice crystal size distributions. {\em J. Atmos. Sci.,}
{\bf 60,} 544--560

\bibitem[Field \etal (2007)]{fcap} Field, P. R., A. J. Heymsfield and C. H. Twohy 2007: Determination of the combined ventilation factor and capacitance for ice crystal aggregates from airborne observations in a tropical anvil cloud. Submitted to {\em J. Atmos. Sci.}

\bibitem[Forbes and Hogan (2006)]{fandhogan} Forbes, R. M. and R. J. Hogan 2006: Observations of the depth of ice particle evaporation beneath frontal cloud to improve NWP modelling. {\em Q. J. R. Meteorol. Soc.,}
{\bf 132,} 865--883

\bibitem[Fukuta and Takahashi (1999)]{fukuta} Fukuta, N. and T. Takahashi 1999: The growth of atmospheric ice crystals: a summary of findings in vertical supercooled cloud tunnel studies. {\em J. Atmos. Sci.,}
{\bf 56,} 1964--1979

\bibitem[Goold \etal (2005)]{goold} Goold, N. R., E. Somfai and R. C. Ball 2005: Anisotropic diffusion-limited aggregation in three dimensions: universality and nonuniversality. {\em Phys. Rev. E,}
{\bf 72,} 031403

\bibitem[Hall and Pruppacher (1976)]{handp} Hall, W. D. and H. R. Pruppacher 1976: The survival of ice particles falling from cirrus clouds in subsaturated air. {\em J. Atmos. Sci.,}
{\bf 33,} 1995--2006

\bibitem[Heymsfield and Iaquinta (2000)]{heyms} Heymsfield, A. J. and J. Iaquinta 2000: Cirrus crystal terminal velocities. {\em J. Atmos. Sci.,}
{\bf 57,} 916--938

\bibitem[Heymsfield \etal (2002)]{heym02} Heymsfield, A. J., S. Lewis, A. Bansemer, J. Iaquinta, L. M. Miloshevich, M. Kajikawa, C. Twohy and M. R. Poellot 2002:
A general approach for deriving the properties of cirrus and stratiform ice particles. {\em J. Atmos.
Sci.}, {\bf 60}, 1795--1808.

\bibitem[Hwang and Mascagni (2004)]{hwang04} Hwang, C.-O. and M. Mascagni 2004: Electrical capacitance of the unit cube. {\em J. Appl. Phys.,}
{\bf 95,} 3798--3802

\bibitem[Jiusto and Weickmann (1973)]{jiusto} Jiusto, J. E. and H. K. Weickmann 1973: Types of snowfall. {\em Bull. Amer. Met. Soc.,}
{\bf 54,} 1148--1162

\bibitem[Khain and Sednev (1996)]{khain} Khain, A. P. and I. Sednev 1996: Simulation of precipitation formation in the Eastern Mediterranean coastal zone using a spectral microphysics cloud ensemble model {\em Atmos. Res.,}
{\bf 43,} 77--110

\bibitem[Korolev and Isaac (2004)]{kandi04} Korolev, A. and G. A. Isaac 2004: Observations of sublimating ice particles in clouds. {\em Proceedings of the $14^{th}$ International Conference on Clouds and Precipitation,} 808--811

\bibitem[Korolev \etal (2004)]{kor04} Korolev, A., M. P. Bailey, J. Hallett and G. A. Isaac 2004: Laboratory and in-situ observations of deposition growth of frozen drops. {\em J. Appl. Met.,} {\bf 43} 612--622

\bibitem[Langer (1980)]{langer} Langer, J. S. 1980: Instabilities and pattern formation in crystal growth. {\em Rev. Mod. Phys.,}
{\bf 52,} 1--28

\bibitem[Liu \etal (2003)]{liu} Liu, H.-C., P. K. Wang and R. E. Schlesinger 2003: A numerical study of cirrus clouds. Part I: Model description. {\em J. Atmos. Sci.,}
{\bf 60,} 1075--1084

\bibitem[Macke \etal (1996)]{macke} Macke, A., J. Mueller and E. Raschke 1996: Single scattering properties of atmospheric ice crystals. {\em J. Atmos. Sci.,}
{\bf 53,} 2813--2825

\bibitem[Magee \etal (2006)]{magee} Magee, N., A. M. Moyle and D. Lamb 2006: Experimental determination of the deposition coefficient of small cirrus-like ice crystals near $-50^{\circ}$C. {\em Geophys. Res. Lett.,}
{\bf 33,} L17813

\bibitem[Mascagni and Hwang (2003)]{mas03} Mascagni, M. and C.-O. Hwang 2003: $\epsilon$-shell error analysis for `walk on spheres' algorithm. {\em Math. \& Comp. in Simul.,}
{\bf 63,} 93--104

\bibitem[McDonald (1963)]{mcd} McDonald, J. E. 1963: Use of the electrostatic analogy in studies of ice crystal growth. {\em Z. Angew. Math. Phys.,}
{\bf 14,} 610--620

\bibitem[Mullins and Sekerka (1963)]{ms63} Mullins, W. W. and R. F. Sekerka 1963: Morphological stability of a particle growing by diffusion or heat flow. {\em J. Appl. Phys.,}
{\bf 34,} 323--329

\bibitem[Nelson (1998)]{nelson} Nelson, J. 1998: Sublimation of ice crystals. {\em J. Atmos. Sci.,}
{\bf 55,} 910--919

\bibitem[Podzimek (1966)]{podzimek} Podzimek, J. 1966: Experimental determination of the capacity of ice crystals. {\em Studia Geophys. Geodet.,}
{\bf 10,} 235--238

\bibitem[P\'{o}lya and Szeg\~{o} (1951)]{polya}
P\'{o}lya and Szeg\~{o} 1951,
\textit{Isoperimetric inequalities in mathematical physics: Annals of mathematical studies}, Princeton University Press, Princeton.

\bibitem[Pruppacher and Klett (1997)]{ppk}
Pruppacher H. R. and J. D. Klett 1997,
\textit{Microphysics of clouds and precipitation}, Springer, New York.

\bibitem[Sander (2000)]{sander} Sander, L. M. 2000: Diffusion-limited aggregation: a kinetic critical phenomenon. {\em Contemporary Phys.,}
{\bf 41,} 203--218

\bibitem[Smythe (1962)]{smythe} Smythe, W. R. 1962: Charged right circular cylinder. {\em J. Appl. Phys.,}
{\bf 33,} 2966--2967

\bibitem[Swanson \etal (1999)]{swan99} Swanson, B. D., N. J. Bacon, E. J. Davis and M. B. Baker 1999: Electrodynamic trapping and manipulation of ice crystals. {\em Q. J. R. Meteorol. Soc.,}
{\bf 125,} 1039--1058

\bibitem[Westbrook \etal (2004)]{west4a} Westbrook, C. D., R. C. Ball, P. R. Field and A. J. Heymsfield 2004: Universality in snowflake aggregation. {\em Geophys. Res. Lett.,}
{\bf 31,} L15104--15107

\bibitem[Westbrook \etal (2006)]{west6} Westbrook, C. D., R. J. Hogan, A. J. Illingworth and E. J. O'Connor 2006: Theory and observations of ice particle evolution in cirrus using Doppler radar: evidence for aggregation. {\em Geophys. Res. Lett.,} {\bf 34,} L02824

\bibitem[Wilson and Ballard (1999)]{wilson} Wilson, D. R. and S. P. Ballard 1999: A microphysically based precipitation scheme for the UK Meterological Office Unified Model, {\em Q. J. R. Meteorol. Soc.,}
{\bf 125,} 1607--1636

\bibitem[Wintle (2004)]{wint} Wintle, H. J. 2004: The capacitance of the cube and square plate by random walk methods, {\em J. Electrostatics,}
{\bf 62,} 51--62

\bibitem[Witten and Sander (1981)]{wit} Witten, T. A. and L. M. Sander 1981: Diffusion limited aggregation, a kinetic critical phenomenon, {\em Phys. Rev. Lett.,}
{\bf 47,} 1400--1403

\bibitem[Woods \etal (2006)]{woods} Woods, C. P., M. T. Stoelinga, J. D. Locatelli and P. V. Hobbs 2006: The IMPROVE-1 Storm of 1-2 February 2001. Part III: Sensitivity of a mesoscale model simulation to the representation of snow particle types and testing of a bulk microphysical scheme with snow habit prediction. {\em J. Atmos. Sci.,} in press.

\bibitem[Zhou \etal (1994)]{zhou} Zhou, H.-X., A. Szabo, J. F. Douglas and J. B. Hubbard 1994: A Brownian dynamics algorithm for calculating the hydrodynamic friction and the electrostatic capacitance of an arbitrarily shaped object, {\em J. Chem. Phys,}
{\bf 100,} 3821--3826

\end{thebibliography}
\end{document}